\documentclass[cameraready]{Interspeech}

\title{A Fusion-Aware Two-Stage Framework for Mispronunciation Detection and Diagnosis in Low-Resource Modern Standard Arabic}
\author[affiliation={1}, equalcontribution]{Jing}{Yang}
\author[affiliation={1}, equalcontribution]{Shuqing}{Zhang}
\author[affiliation={1}]{Yongyi}{Deng}
\author[affiliation={1}]{Pan}{Li}
\author[affiliation={2}]{Ting}{Dang}
\author[affiliation={1}, correspondingauthor]{Gongping}{Huang}
\author[affiliation={3}]{Jingdong}{Chen}
\author[affiliation={4}]{Jacob}{Benesty}

\address{
    $^1$ School of Electronic Information, Wuhan University, Wuhan Hubei, 430072, China\\ 
    $^2$ School of Computing and Information Systems, University of Melbourne, Melbourne, Australia \\
    $^3$ Northwestern Polytechnical University \\
    $^4$ University of Quebec
}

\email{yjing\_whu@whu.edu.cn, 2024302121412@whu.edu.cn
}

\usepackage{comment}
\usepackage{hyperref}  

\begin{document}

\maketitle

\begin{abstract}
Accurate phoneme recognition is pivotal for mispronunciation detection and diagnosis (MDD) in modern standard Arabic (MSA), yet remains constrained by data scarcity and the synthetic-real domain gap. This work proposes a two-stage end-to-end framework. It integrates a pre-trained encoder with causal dilated temporal convolutional networks to preserve fine-grained phonetic variations. A hierarchical two-stage strategy first learns general mappings from native/synthetic corpora, then adapts to scarce real learner data to mitigate domain shift without over-correction. Prediction stability is further enhanced via multi-checkpoint ensemble inference with N-gram rescoring. Evaluated on the QuranMB.v2 test set, our system achieves an F1-score of $0.7201$, a $63.1$\% relative improvement over baseline ($0.4414$). This performance ranks at the top of the IqraEval.2 Challenge, establishing a new state-of-the-art for low-resource MSA in MDD.
\end{abstract}

\section{Introduction}
\label{sec:intro}

Computer-aided pronunciation training (CAPT) has become an essential tool for second language acquisition. Mispronunciation detection and diagnosis (MDD)~\cite{7752846, 9689243, 8682654,10485145,8461841,8462635} serves as a critical component of CAPT systems, and its effectiveness hinges on a specific capability: for fine-grained diagnosis, the ability to distinguish canonical from non-canonical pronunciation units is paramount. Unlike standard automatic speech recognition (ASR), which prioritizes semantic intelligibility and often tends to "correct" learner errors, effective MDD requires the system to faithfully capture these non-canonical variations to enable precise error localization and diagnosis. This distinction is particularly critical for modern standard Arabic (MSA), a language with a complex phonological inventory (e.g., pharyngeal and emphatic consonants) that poses significant challenges for automated detection~\cite{Kheir2025TowardsAU}.

Recent studies have explored deep learning architectures for CAPT, specifically integrating connectionist temporal classification (CTC)~\cite{8461929, LI2020107392, 10.1145/1143844.1143891, 8461935, sak15_interspeech} with self-supervised acoustic encoders~\cite{9585401,NEURIPS2020_92d1e1eb,radford2023robust}. While these end-to-end approaches~\cite{8682654, yan20_interspeech, Alrashoudi2025, wu21h_interspeech,cmc.2023.033457} have demonstrated notable improvements in languages like English, their application to MSA remains constrained. Existing models struggle to capture MSA's subtle articulatory variations, while the scarcity of standardized resources and benchmarks further stifles progress in this domain. Although recent initiatives, such as the Qur'anic mispronunciation benchmark~\cite{Kheir2025TowardsAU}, have begun to bridge this resource gap, critical obstacles persist, such as bridging the distributional gap between synthetic and real learner speech and enabling fine-grained error diagnosis.

To address these challenges, we propose a robust end-to-end MSA phoneme recognition framework specifically tailored for MDD. Our approach substantially outperforms existing methods on the Qur'anic mispronunciation benchmark, driven by three key innovations. First, we design a hybrid architecture that integrates a large-scale multilingual pre-trained encoder with causal dilated temporal convolutional networks (TCNs)~\cite{stoter2019temporal, Haque2019AudiolinguisticEF} in a sequential, end-to-end manner. Unlike transformer-based models~\cite{Alrashoudi2025} that inherently prioritize global semantic coherence, our TCN-based module explicitly leverages a strong local inductive bias to preserve fine-grained phonetic variations essential for distinguishing canonical from non-canonical pronunciations, all while maintaining computational efficiency. Second, to effectively bridge the distribution gap between controlled synthetic errors and noisy real-world learner speech, we introduce a hierarchical two-stage training paradigm. In the first stage, the model learns general acoustic-phonetic mappings from large-scale native and synthetic data; in the second stage, it adapts specifically to the unique distribution of scarce real learner recordings, thereby mitigating domain shift without overfitting to synthetic artifacts. 
Third, we employ a multi-checkpoint ensemble inference method combined with modified Kneser-Ney (MKN) N-gram language model rescoring~\cite{479394}, leveraging complementary features from different training phases to enhance prediction stability on unseen data.

The effectiveness of our proposed framework is decisively validated on the blind QuranMB.v2 test set. Our system achieves a phoneme-level F1-score of 0.7201, surpassing the official baseline ($0.4414$) by a remarkable $63.1$\% relative improvement.
This result ranks at the top of the IqraEval.2 Challenge at Interspeech 2026~\footnote{\href{https://huggingface.co/spaces/IqraEval}{hf.co/spaces/IqraEval}}. It establishes a new state-of-the-art for low-resource Arabic MDD, demonstrating exceptional generalization capability and resilience against data scarcity.

\begin{figure*}[htp]
  \centering
  \includegraphics[width=\linewidth]{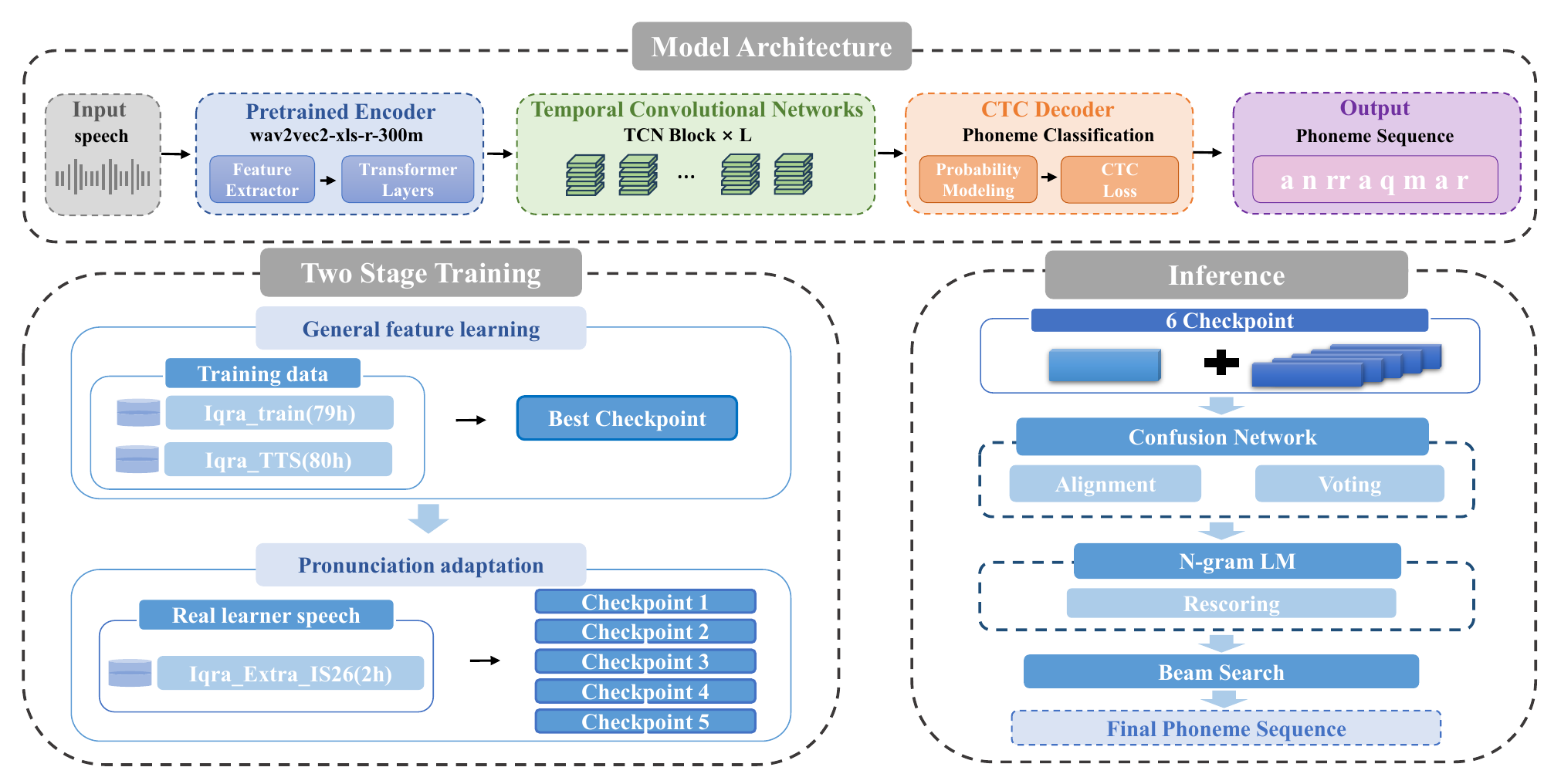} 
  \caption{Overview of the proposed framework. The system consists of a wav2vec2-XLS-R encoder coupled with causal dilated TCNs. Training follows a two-stage hierarchy: (1) general feature learning on native and synthetic data, and (2) pronunciation pattern adaptation on real learner data. Inference employs a multi-checkpoint ensemble fused via CN and rescored with N-gram language model.}
  \label{fig:framework}
\end{figure*}

Our main contributions are summarized as follows:
\begin{itemize}
    \item \textbf{A Robust End-to-End Architecture Tailored for MSA:} We propose a hybrid framework that integrates a large-scale multilingual pre-trained encoder with causal dilated TCNs. The design is optimized to preserve fine-grained phonetic details that are critical for the complex phoneme inventory of MSA. In doing so, it reduces the tendency to over-smooth subtle acoustic anomalies that is often observed in recurrent architectures such as LSTMs, and it outperforms LSTM-based baselines in low-resource settings.
    
    \item \textbf{Hierarchical Two-Stage Training Strategy:} We propose a novel training paradigm that sequentially optimizes for general feature learning (on native+synthetic data) and pronunciation pattern adaptation (on real learner data). This strategy effectively resolves the synthetic-real domain gap identified in the IqraEval benchmark.

    \item \textbf{Diversity-Aware Ensemble Inference:} We introduce a multi-checkpoint ensemble inference framework that fuses predictions from multiple model states via confusion networks (CN) and enhances stability through MKN N-gram language model rescoring. This approach leverages complementary strengths across training phases, achieving a $5.5$\% relative improvement over the single-checkpoint baseline.
\end{itemize}

The remainder of this paper is organized as follows: Section~\ref{sec:method} details the proposed methodology. Section~\ref{sec:experiments} presents the experimental setup and results. Finally, Section~\ref{sec:conclusion} concludes the paper.

\section{Proposed Method}
\label{sec:method}
As illustrated in Figure~\ref{fig:framework}, our approach comprises three core components: (1) a hybrid model architecture integrating a large-scale multilingual pre-trained encoder with causal dilated TCNs; (2) a hierarchical two-stage training strategy that first learns general acoustic-phonetic representations from native and synthetic data, then adapts the model to the distribution of real learner speech to capture non-canonical pronunciation patterns; and (3) a diversity-aware ensemble inference mechanism leveraging CN and linguistic constraints.

\subsection{Hybrid Model Architecture}

Our model uses a ''Pre-trained Encoder + causal dilated TCNs + CTC Decoding'' framework. The encoder incorporates cross-lingual and language-specific acoustic priors. The TCNs model local temporal patterns via causal dilated convolutions. The CTC loss enables end-to-end training without explicit frame-level alignment. 

\subsubsection{Multilingual Pre-trained Encoder}
To overcome the limitations of low-resource MSA data, we utilize wav2vec2-XLS-R~\cite{NEURIPS2020_92d1e1eb} as the upstream acoustic encoder. Pre-trained on 126 languages with billions of audio samples, this model possesses rich cross-lingual acoustic priors that provide a robust foundation for representing diverse phonetic units. Crucially, its exposure to Arabic during pre-training endows it with the latent capacity to distinguish fine-grained phonetic features inherent to MSA, such as emphatic consonants ($/S/$ vs. $/s/$), pharyngeals, and vowel. During the training phase, we employ a low learning rate fine-tuning strategy for the encoder weights. This approach leverages the generalized acoustic knowledge acquired during pre-training to facilitate rapid adaptation to the specific MSA phonetic inventory, thereby mitigating catastrophic forgetting while effectively specializing the model's latent capabilities for precise phoneme discrimination.

\subsubsection{Causal Dilated Temporal Convolutional Networks}
For downstream temporal modeling, we employ causal dilated TCNs~\cite{stoter2019temporal, Haque2019AudiolinguisticEF}, a choice driven by the specific requirements of MDD in MSA. Since the complex inventory is characterized by subtle emphatic contrasts and geminated consonants, our primary objective shifts from maximizing global semantic coherence to preserving local acoustic fidelity. To achieve this, we utilize causal dilated TCNs to capture short-term articulatory cues without reliance on long-range semantic context:
\begin{itemize}
    \item \textbf{Local Feature Sensitivity}: The inherent local inductive bias of convolutions forces the model to focus on short-term acoustic patterns critical for distinguishing subtle MSA contrasts (e.g., emphatic $/S/$ vs. non-emphatic $/s/$). This prevents the over-reliance on broad context that often masks pronunciation errors in attention-based models.
    
    \item \textbf{High-Resolution Context Modeling}: Dilated convolutions exponentially expand the receptive field to capture prosodic dependencies (e.g., the duration cues essential for gemination) without sacrificing the frame-level temporal resolution required for precise error localization.
    \item \textbf{Strict Causality}: The causal structure restricts the receptive field to past and current inputs, avoiding reliance on future long-range context. This design helps prevent fine-grained non-canonical pronunciation patterns from being over-smoothed by broader contextual aggregation, thereby preserving subtle local acoustic anomalies.
\end{itemize}
Leveraging the inherent local inductive bias of convolutions, this architecture explicitly prioritizes local discriminative power over the global semantic smoothing typical of attention mechanisms, thereby significantly enhancing diagnostic precision for the challenging phonetic landscape of MSA.

\subsection{CTC Objective}
\label{subsec:ctc_objective}

We employ the CTC loss function~\cite{8461929} to align variable-length acoustic frames with target phoneme sequences. This approach is particularly effective for handling the diverse durations of MSA phonemes, including geminated consonants (e.g., /bb/), as it eliminates the need for explicit frame-level alignments. By introducing a dedicated blank symbol, CTC marginalizes over all possible monotonic alignments between the input and output sequences, thereby enabling end-to-end training directly from unsegmented data. During inference, we utilize beam search decoding integrated with language model rescoring and multi-checkpoint fusion to maximize diagnostic accuracy, rather than relying on simple greedy decoding.

\subsection{Hierarchical Two-Stage Training Strategy}

To effectively bridge the distribution gap between controlled synthetic errors and noisy real-world pronunciations, we propose a hierarchical two-stage training protocol that decouples general acoustic-phonetic learning from domain-specific adaptation. This strategy addresses the scarcity of annotated real learner data by leveraging large-scale synthetic resources for initialization before fine-tuning on authentic samples.

\textbf{Stage 1: General Feature Learning .} 
The primary objective of this stage is to establish a robust mapping between MSA acoustics and the standard phoneme inventory. We train the model on a joint corpus comprising high-quality native speech and large-scale synthetic data with systematically injected confusion errors. The native component provides canonical pronunciation paradigms, while the synthetic component augments the diversity of error patterns beyond what is typically available in wild-collected datasets. By optimizing on this combined corpus, the model learns fundamental acoustic-phonetic correlations and generalizes across a wide spectrum of both canonical and simulated non-canonical pronunciations. This phase serves as a crucial initialization step, equipping the model with broad acoustic priors that mitigate the risk of overfitting in subsequent low-resource fine-tuning.

\textbf{Stage 2: Pronunciation  Adaptation.} 
Building upon the weights optimized in Stage 1, the second stage focuses exclusively on adapting the model to the specific distribution of real learner speech. We fine-tune the model on a scarce but critical dataset of authentic learner recordings, which naturally contain uncontrolled disfluencies, dialectal interferences, and complex error types that are difficult to simulate synthetically. This phase shifts the model's decision boundaries to accommodate the noisy characteristics of wild data, effectively closing the synthetic-real gap. Crucially, instead of retaining only the final converged model, we preserve multiple checkpoints from different convergence stages during this phase. These checkpoints capture diverse states of the model's manifold, reflecting varying degrees of sensitivity to non-native features. This collection forms a diverse ensemble pool, enhancing robustness against overfitting and improving generalization in the final inference stage.

\subsection{Diversity-Aware Ensemble Inference}
To maximize recognition stability and accuracy, we introduce a post-processing pipeline that fuses predictions from multiple model states and incorporates linguistic constraints, moving beyond simple single-model greedy decoding.

\subsubsection{Confusion Network Construction}
Let $\mathcal{M} = \{m_1, m_2, \dots, m_K\}$ denote our ensemble pool, consisting of the optimal checkpoint from Stage~1 and a selected subset of checkpoints from Stage~2. For a given input utterance, each model $m_k$ generates a hypothesis sequence $H_k$. 
These $K$ hypotheses are aligned using a weighted Levenshtein distance algorithm to construct a CN, represented as a linear chain of slots. Each slot contains a set of candidate phonemes and their posterior probabilities, which are derived from their normalized occurrence frequencies and associated confidence scores across the aligned hypotheses. 
This ensemble structure effectively captures diverse prediction patterns inherent in different training stages. 
Finally, probability sharpening is applied to accentuate high-confidence candidates, while pruning eliminates low-probability paths, yielding a compact and representative search space for final decoding.


\subsubsection{N-gram Language Model Rescoring}
To ensure phonotactic consistency and correct locally plausible but globally inconsistent hypotheses, we introduce a Kneser–Ney smoothed N-gram language model~\cite{479394} for rescoring candidate sequences within the CN. Crucially, it is not a static external model pre-trained on reference corpora. Instead, it acts as a self-induced prior, estimated directly from pooled hypotheses generated by multiple checkpoints for the current utterance. By aggregating these predictions, it captures robust phonotactic patterns to regularize path selection toward common N-gram patterns supported by checkpoint outputs, thereby stabilizing multi-checkpoint aggregation. Since it relies solely on the fused hypotheses, no reference labels are required. This model employs absolute discounting and back-off interpolation to alleviate data sparsity, providing a dynamic sequence prior that complements acoustic posteriors and enhances decoding robustness.


\subsubsection{Beam Search}
The final phoneme sequence $\hat{Y}$ is obtained by finding the candidate sequence $Y$ that maximizes the combined score:
\begin{equation}
    \hat{Y} = \underset{Y}{\arg\max} \left( \log P_{\text{acoustic}}(Y) + \lambda \cdot \log P_{\text{LM}}(Y) \right),
\end{equation}
where $Y$ denotes any candidate phoneme sequence generated by CN, $\hat{Y}$ represents the sequence with the highest combined score, and $\lambda$ is a hyperparameter balancing acoustic and linguistic contributions. In our experiments, $\lambda$ is set to $0.2$, prioritizing acoustic evidence to ensure that non-canonical pronunciation patterns (errors) are faithfully preserved rather than being overwritten by linguistic priors. This joint decoding process ensures that the output is not only acoustically probable across multiple model states but also linguistically coherent within the context of MSA.

\section{Experiments}
\label{sec:experiments}

\subsection{Experimental Setup}
\label{ssec:setup}
\textbf{Datasets.} We utilize the multi-source corpus from the IqraEval challenge~\footnote{ \href{https://huggingface.co/spaces/IqraEval}{hf.co/spaces/IqraEval}}, partitioned for our two-stage strategy:
(1) Training: Stage~1 uses $\sim$79h of native MSA (Iqra\_train~\footnote{ \href{https://huggingface.co/datasets/IqraEval/Iqra_train}{hf.co/datasets/IqraEval/Iqra\_train}}) 
and $\sim$80h of synthetic error-injected speech (Iqra\_TTS~\footnote{ \href{https://huggingface.co/datasets/IqraEval/Iqra_TTS}{hf.co/datasets/IqraEval/Iqra\_TTS}}). 
Stage~2 adapts on $\sim$2h of real learner recordings (Iqra\_Extra\_IS26~\footnote{ \href{https://huggingface.co/datasets/IqraEval/Iqra_Extra_IS26}{hf.co/datasets/IqraEval/Iqra\_Extra\_IS26}}) to capture natural disfluencies.
(2) Validation: A 3.4h held-out subset of Iqra\_train guides checkpoint selection.
(3) Test: Evaluation is conducted on the blind QuranMB.v2 benchmark, ensuring fair comparison with baselines on annotated real learner speech.

\textbf{Implementation Details.} Implemented via the s3prl toolkit, our model initializes the encoder with wav2vec2-XLS-R-300m~\cite{NEURIPS2020_92d1e1eb}. Experiments are conducted on a single NVIDIA RTX 4090 GPU. We employ Adam optimization with differential learning rates ($1\times10^{-5}$ for the encoder, $1\times10^{-4}$ for the TCNs head) and a batch size of 4 with mixed-precision (FP16). For the final ensemble inference, we construct a pool of $K=6$ models, consisting of the single best-performing checkpoint from Stage~1 and five checkpoints randomly sampled from the top-performing candidates of Stage~2 (ranked by validation F1-score) to enhance prediction diversity. Finally, a 3-gram language model is employed for sequence rescoring to refine the output hypothesis and improve contextual coherence.

\begin{table}[t]
    \centering
    \footnotesize 
    \setlength{\tabcolsep}{4pt} 
    \renewcommand{\arraystretch}{1.1} 
    \renewcommand{\arraystretch}{1.2}
    \caption{Performance comparison on the QuranMB.v2 blind test set. ``Rel. Imp.'' denotes the relative improvement over the baseline system. 
    }
    \label{tab:results}
    \begin{tabular*}{\columnwidth}{@{\extracolsep{\fill}} lcc}
        \toprule
        \textbf{System} & \textbf{F1-Score} & \textbf{Rel. Imp. (\%)} \\
        \hline
        Baseline & $0.4414$ & - \\
        \textbf{Ours (Two-Stage + Ensemble)} & \textbf{$0.7201$} & \textbf{$+63.1$} \\
        \midrule
        \textit{Two-Stage (Single CKPT)} & $0.6825$ & $+54.6$ \\
        Stage 1 Only (Single CKPT) & $0.4629$ & $+4.9$ \\
        Stage 2 Only (Single CKPT) &  $0.6681$ & $+51.4$ \\
        Mix (Single CKPT) &  $0.4305$ & $-2.5$ \\
        \midrule
        Stage 2 Only (LSTM) &  $0.6467$ & $+46.5$ \\
        Stage 2 Only (Transformer) &  $0.6000$ & $+35.9$ \\
        \bottomrule
    \end{tabular*}
\end{table}

\subsection{Results}
\label{ssec:results}

Table~\ref{tab:results} presents the performance on the blind QuranMB.v2 test set. Our framework achieves a phoneme-level F1-score of $0.7201$, a $63.1$\% relative improvement over the official baseline ($0.4414$). This gain demonstrates the effectiveness of our approach in bridging the synthetic-real domain gap and capturing fine-grained phonetic variations, even under low-resource conditions. Moreover, strong performance on unseen data confirms the robust generalization of our ensemble mechanism.

\subsection{Analysis and Discussion}
\label{ssec:discussion}

To dissect the contributions of our hierarchical strategy and architectural choices, we conducted comprehensive ablation experiments on the QuranMB.v2 test set. Table~\ref{tab:results} summarizes the results, yielding three key insights:

\begin{enumerate}[leftmargin=*]
    \item \textbf{Insufficiency of Synthetic Data \& Mixing}: 
    Training solely on Stage~1 data yields a marginal $+4.9\%$ gain, confirming the insufficiency of synthetic data. Worse, naive mixing (Mix) underperforms the baseline ($-2.5\%$), indicating that unaligned aggregation exacerbates domain shift. This validates our sequential strategy over simple augmentation.
    
    \item \textbf{Synergy of Two-Stage Adaptation}: 
    While training exclusively on scarce real data yields an F1-score of $0.6681$, our two-stage strategy further elevates it to $0.6825$. This confirms that Stage $1$ initialization effectively prevents overfitting and improves generalization on scarce real data.
    
    \item \textbf{Architectural Superiority}: Our causal dilated TCNs backbone proves highly effective for this task. As shown in Table~\ref{tab:results}, the TCN-based Stage 2 model substantially outperforms both LSTM and Transformer~\cite{NIPS2017_3f5ee243} variants. This is largely attributable to the TCN’s superior inductive bias for modeling local phonetic dependencies, which enables more accurate representation learning in low-resource settings compared to attention-based or recurrent architectures.
\end{enumerate}

\textbf{Impact of Inference Optimization.}
Beyond the core architecture, our inference strategies provide substantial gains. The integration of multi-checkpoint ensemble and N-gram language model rescoring boosts the F1-score from $0.6825$ to $0.7201$, delivering an additional $5.5$\% relative improvement. This confirms that leveraging prediction diversity and phonotactic constraints is crucial for maximizing performance in low-resource MDD systems.

\section{Conclusion}
\label{sec:conclusion}
We presented a robust end-to-end framework for modern standard Arabic MDD, addressing data scarcity and domain shift via three innovations: causal dilated TCNs for fine-grained feature preservation, a hierarchical two-stage training strategy to bridge synthetic-real gaps, and ensemble inference for stability. Evaluated on the blind QuranMB.v2 test set, our system achieved an F1-score of $0.7201$ ($+63.1\%$ over baseline), ranking at the top of the IqraEval.2 Challenge. Ablation studies confirmed that the core gain ($+54.6\%$) stems from the synergy of our architecture capturing fine-grained phonetic nuances and our sequential adaptation strategy bridging the domain gap without overfitting; additionally, multi-checkpoint fusion and linguistic rescoring provided a crucial boost ($+5.5\%$), maximizing stability on unseen data. This work establishes a new state-of-the-art for low-resource MSA MDD and offers a scalable blueprint for pronunciation assessment in other data-scarce languages.

\section{Acknowledgments}
This work was supported by the National Natural Science Foundation (NSFC) of China under Grant 62471340. The authors would like to thank the Supercomputing Center of Wuhan University for providing the computational resources.

\section{Generative AI Use Disclosure}
The authors declare that generative AI tools were used solely for editing, polishing, and improving the grammar and readability of the manuscript. No generative AI tool was used to produce a significant part of the manuscript, and no AI tool is listed as a co-author.
\bibliographystyle{IEEEtran}
\bibliography{mybib}

\end{document}